\documentclass{iopart}
\usepackage[OT2,OT1]{fontenc}
\def\cyr{\fontencoding{OT2}\fontfamily{wncyr}\selectfont}

\newcommand{\lesssim}{\raisebox{0.3mm}{\em $\, <$} 
\hspace{-3.3mm} \raisebox{-1.8mm}{\em $\sim \,$}}
\newcommand{\gtrsim}{\raisebox{0.3mm}{\em $\, >$} 
\hspace{-3.3mm} \raisebox{-1.8mm}{\em $\sim \,$}}
\begin{document}

\title{Summary of Working Group 2 }

\author{Bruno Autin$*$, Deborah A. Harris\dag
\footnote[3]{Corresponding author (dharris@fnal.gov)}, 
Steve F. King\ddag,\\
Kevin S. McFarland$**$ and Osamu Yasuda\P\
}
\address{* CERN, Geneva, Switzerland} 
\address{** Department of Physics and Astronomy, 
University of Rochester, Rochester, NY 14627, USA } 
\address{\dag\ Fermi National Accelerator Laboratory,
 P.O. Box 500, Batavia, IL 60510, USA}

\address{\ddag\ Department of Physics and Astronomy, \\ 
        University of Southampton, Southampton SO17 1BJ, U.K.}

\address{\P\ Department of Physics,
Tokyo Metropolitan University, \\
Minami-Osawa, Hachioji, Tokyo 192-0397, Japan}

\begin{abstract}
Issues on the physics of, beamlines for, and detectors of 
neutrino oscillation
discussed in Working Group 2 at Nufact'02 are summarized.
\end{abstract}




\section{Introduction} 

     The way people are thinking about neutrino oscillations and how 
a neutrino factory fits into a program of measurements has changed 
dramatically between previous NuFact workshops and this one.  
Although we still consider a neutrino factory the ultimate facility 
to do precision oscillation measurements, we are becoming more 
aware of what 
one can learn with conventional neutrino beams.  Understanding how much
of the possible parameter space one can explore with conventional 
beams, and how these experiments might complement neutrino factory 
measurements  has kept many phenomenologists busy this past year.  
The oscillation working group also has included for the first time discussions 
on different proposals for these super conventional neutrino beams, as well 
as a new idea for making a pure $\nu_e$ or $\bar\nu_e$ 
beam, which can be done by 
accelerating radioactive ions which undergo beta decay.  With these 
new techniques for measurements, new issues become relevant:  
cross sections, beam systematics, and backgrounds, 
the three things that were so straightforward to tackle in the case of 
neutrino factory experiments. 

While the experimental picture is changing, so too is the theoretical
motivation for studying neutrino oscillations.  With the 
concept of extra dimensions becoming more fleshed out in the past 
year, neutrinos have surfaced as an example of particles which have 
significant activity outside the brane we live on. Furthermore, given that 
the SNO Experiment has confirmed solar oscillations, and in particular, 
confirmed that the mixing angle describing solar neutrino oscillations 
is large, the prospect of seeing leptonic  
CP violation in the laboratory is still alive.  
The see-saw mechanism remains an attractive candidate for accounting
for the smallness of neutrino mass and can easily accommodate the
observed large solar angle. Furthermore the see-saw mechanism
allows the possibility that baryon asymmetry is generated in the early
universe, but only if CP is violated in the lepton sector.
In the past year there has been much theoretical work
trying to make the link between the CP violation required by
baryon asymmetry and that potentially
observable in low energy neutrino experiments.

In this summary we try to give an overview of what the issues are in 
both understanding the origin of neutrino oscillations, how to go 
from experimental measurements to oscillation parameters, and finally 
how to make the experimental measurements themselves.  
 
\section{Theory}

Perhaps the zeroth order question is whether we are actually seeing
atmospheric and solar neutrino oscillations, or some other
non-standard effect? Valle \cite{valle} showed that while the case
for atmospheric oscillations looks convincing, there are alternatives
to solar oscillations which fit the data well, such as
spin flavor precession or flavor changing neutrino interactions.
However the magnetic moments and fields, or flavor changing couplings
do not appear very natural, and an oscillation signal from KamLAND
would rule out such interpretations.

Assuming we are seeing oscillations, and hence neutrino mass, then,
ignoring LSND for the moment (discussed later), the atmospheric and solar data
imply two possible patterns of neutrino mass, distinguished
by the sign of $\Delta m_{32}^2$, together with two large mixing angles
and one small mixing angle (also discussed later). From a theoretical
point of view the challenge is to account for such a neutrino spectrum.
The first point to make is that neutrino masses are zero in the Standard
Model, so neutrino oscillations are the first solid evidence for new physics
beyond the Standard Model. Three possible ways to extend the Standard Model
in order to account for the neutrino mass spectrum above
were discussed in this working group, namely the see-saw mechanism,
extra dimensions, and R-parity violating supersymmetry.

In his plenary talk Ross \cite{ross} reviewed the see-saw mechanism
in which super-heavy right-handed Majorana neutrinos are introduced
into the Standard Model, which induce light effective left-handed
Majorana neutrino masses according to the formula $m_{\nu}\sim m_{f}^2/M$
where $m_{f}$ represents a typical quark or charged lepton mass and $M$
is the heavy right-handed neutrino mass scale. He showed that if one
of the right-handed neutrinos plays the dominant role in the see-saw
mechanism then a mass hierarchy and large mixing angles may naturally emerge.

Blazek \cite{blazek} applied the see-saw mechanism to GUT models
such as $SO(10)$ where three families of right-handed neutrinos 
with Majorana masses not far below the GUT scale are predicted.
GUT models require supersymmetry (SUSY) in order to 
allow the gauge couplings to meet, and also to stabilize the hierarchy.
One of the consequences of having a see-saw mechanism with SUSY is
lepton flavor violation (LFV) which was discussed extensively 
in this working group. Blazek \cite{blazek} 
showed that sizable branching ratios for 
$\tau \rightarrow \mu \gamma$ are expected.
Shimizu \cite{shimizu} confirmed this and presented results
for $\mu \rightarrow e \gamma$ which also showed large branching ratios. 
Shimoyama \cite{shimoyama}
showed that in a simplified model where right-handed neutrinos
have universal masses the standard SUSY model implies branching ratios for
$\mu \rightarrow e \gamma$ which exceed the experimental limit.
Finally Koike \cite{koike} studied the LFV process 
$\mu+ N^A_Z\rightarrow e+N^A_Z$ and concluded that $Z\sim 30-60$
represents the optimal range for experimental searches. He also 
showed how it may be possible to distinguish models of new physics
by measurements on several different kinds of nuclei.

The see-saw mechanism also opens up the possibility of generating the 
baryon asymmetry of the universe via leptogenesis.
The idea due to Fukugita and Yanagida \cite{Fukugita:1986hr}
is that in the early universe the super-heavy lepton number violating 
Majorana right-handed neutrinos would have been produced,
and, if CP is violated, could then decay out of equilibrium 
giving a lepton asymmetry which is subsequently converted
into baryon asymmetry via electroweak sphalerons. 
The possible link between the CP violating phases necessary 
for leptogenesis and the CP violating phases measurable by 
low energy neutrino physics was also explored in this working group.
Working in a simple framework of two right-handed neutrinos,
Morozumi \cite{morozumi} discussed the possible correlation
between leptogenesis and low energy phases. Ibarra \cite{ibarra}
discussed the case of three right-handed neutrinos, and showed that
although in general there is no direct link, the low energy observable
phases may contribute to leptogenesis over much of the parameter space.

Neutrino masses in the framework of extra dimensions were
considered by Shafi \cite{shafi}, who specialized to the case of 
a 5D standard model with warped geometry. 
The basic idea is that the Standard Model fermions reside in the 5D bulk,
and the quark and lepton mass hierarchies can be interpreted in a geometrical
way, with the higgs field localized at the TeV brane, and the fermion
masses being in direct proportion to the overlap of their wavefunctions
with the higgs field. In such a set-up
the see-saw mechanism cannot be implemented because of the warped geometry, 
however the desired Majorana masses may arise from non-renormalizable
operators, and the masses again depend on the position of the left-handed
leptons in the bulk.

Neutrino masses in R-parity violating (RPV) SUSY were discussed
by Valle \cite{valle}. In the simplest bilinear RPV scheme one
adds an explicit mass term coupling lepton doublets to a Higgsino
doublet. The left-handed neutrinos are then mixed in the neutralino
mass matrix, resulting in a single neutrino Majorana mass term.
The other neutrino masses and mixings are generated by loop corrections,
depending on the RPV SUSY parameters. This results in a testable relation 
between neutrino masses and RPV SUSY physics at colliders.

\section{Phenomenology}

\subsection{Three flavor mixing}

The observation of atmospheric neutrinos (See, e.g.,
Ref.\cite{Kajita:2001mr})
and solar neutrinos (See, e.g., Ref.\cite{Bahcall:2000kh}) implies 
that not only are neutrinos massive, but that the mass eigenstates
($\nu_1,\nu_2,\nu_3$) with mass ($m_1,m_2,m_3$)
are not identical to the flavor eigenstates 
($\nu_e, \nu_\mu, \nu_\tau$).  The unitary matrix which translates
from the mass to the flavor basis is usually parameterized by three 
mixing angles ($\theta_{12},\theta_{23},\theta_{13}$), 
and a CP violating phase ($\delta$), and is often referred to as the 
MNSP \cite{Maki:1962mu,Pontecorvo:1968fh} matrix, after Maki, Nakagawa, 
Sakata, and Pontecorvo.  The atmospheric data largely
constrain ($|\Delta m_{32}^2|$, $\theta_{23}$) and the solar
data constrain ($\Delta m_{21}^2$, $\theta_{12}$) in this three flavor
framework of neutrino oscillations\cite{Hagiwara:pw},
where $\Delta m^2_{ij}\equiv m^2_i-m^2_j$.
The constraint from the CHOOZ
result \cite{Apollonio:1999ae} implies $\sin^22\theta_{13}\lesssim 0.1$.
So in vacuum the MNSP matrix looks roughly like
\begin{eqnarray}
U_{MNSP} \simeq
\left(
\begin{array}{ccc}
c_\odot & s_\odot &  \epsilon\\
-s_\odot/\sqrt{2} &
c_\odot/\sqrt{2} & 1/\sqrt{2}\\
s_\odot/\sqrt{2} &
-c_\odot/\sqrt{2} & 1/\sqrt{2}\\
\end{array}
\right),\nonumber
\end{eqnarray}
where we have used $\theta_{23}\simeq \pi/4$
and $|\epsilon|\ll 1$, and $c_\odot=\cos\theta_{12}$, and 
$s_\odot=\sin\theta_{12}$.

Choubey \cite{choubey} performed model independent and model dependent 
analyses of solar neutrino data including the neutral current event rate 
from SNO \cite{Ahmad:2002ka}.
She concluded that the Large Mixing Angle solution gives the best fit
($\sin^22\theta_{12}\simeq 0.8$)
and the LOW solution is allowed only at the $3\sigma$ level. 
Valle \cite{valle} also gave results on his solar neutrino analyses
and his results agree with those in \cite{choubey}.
While whether the LMA solution is the correct one or not will be
soon confirmed by the KamLAND experiment, it should be
noticed that the goodness of fit (GOF) for each of the other solar solutions
(LOW, Quasi Vacuum Oscillation, Small Mixing Angle solutions)
is still acceptable.  Namely, according to \cite{choubey},
we have $\chi^2$(LMA)=40.57, $\chi^2$(LOW)=50.62, $\chi^2$(QVO)=56.11, and 
$\chi^2$(SMA)=70.97 for 45 degrees of freedom.  
This implies that LMA is 65.99\% (0.44$\sigma$CL) likely,
LOW is 26.14\% (1.12$\sigma$CL) likely,
QVO is 12.39\% (1.54$\sigma$CL) likely, and
SMA is 0.81\% (2.65$\sigma$CL) likely.
As far as GOF is concerned, none of these solutions is
excluded from a conservative point of view.\footnote{The
criterion in the 1996 version of
Particle Data Book (sect. 28.5 in \cite{Barnett:1996hr})
is that 4$\sigma$ or 5$\sigma$ is necessary to exclude
a hypothesis.}
They are disfavored when they are compared with
the LMA solution, i.e.,
$\Delta\chi^2$ = $\chi^2(QVO)-\chi^2$(LMA) = 15.54 (3.53\,$\sigma$CL for
d.o.f. = 2),
$\Delta\chi^2$ = $\chi^2(SMA)-\chi^2$(LMA) = 30.40 (5.16\,$\sigma$CL for
d.o.f. = 2).
On the other hand, \cite{Fogli:2002pt} argues that
GOF is a necessary but not sufficient
criterion for a model to survive.
\cite{Fogli:2002pt} claims that the SMA solution is rejected
because it creates strong tension between not only total rate versus
spectrum data, but also between SNO data versus all the others,
while the LOW and QVO solutions are still acceptable because the
tension is much smaller for them than that for SMA.

With the mass hierarchy $|\Delta m_{21}^2|\ll|\Delta m_{32}^2|$
there are two possible
mass patterns,
depending on whether $\Delta m_{32}^2$ is positive
(normal hierarchy: $m^2_1, m^2_2\ll m^2_3$) or
negative (inverse  hierarchy: $m^2_3\ll m^2_1, m^2_2$).
The quantities one has to determine are $\theta_{13}$,
the sign of $\Delta m_{32}^2$ and the CP phase $\delta$.
To discuss CP violation in long baseline experiments,
it is helpful to have the exact analytical expressions for
the oscillation probabilities in matter.
Kimura \cite{kimura} gave the expressions in a nice way.
The point of his work is that the quantity
$\tilde{X}^{\alpha\beta}_j\equiv\tilde{U}_{\alpha j}\tilde{U}^\ast_{\beta j}$
($\alpha, \beta=e, \mu, \tau$ are the flavor indices,
$j$=1,2,3 is the index of the mass eigen states, and
no sum on $j$ is implied) in matter with constant density can be expressed
linearly in the quantity
$X^{\alpha\beta}_j\equiv U_{\alpha j}U^\ast_{\beta j}$
in vacuum with some coefficients which are functions of
the eigenvalues $\tilde{E}_j$ of the $3\times3$ matrix
$U$diag($E_1,E_2,E_3$)$U^{-1}$+diag($\sqrt{2}G_FN_e$,0,0),
where $E_j\equiv\sqrt{p^2+m^2_j}$.
His results are more useful than previous works
(e.g., \cite{Zaglauer:gz}) since the expression
does not use a particular parametrization for $U_{\alpha j}$
and it is relatively easy to see the condition in
which enhancement of the oscillation probability occurs.
Once we know the form for $\tilde{X}^{\alpha\beta}_j$,
we can write down the the oscillation probability as
\begin{eqnarray}
P(\nu_\alpha\rightarrow\nu_\beta)
=\delta_{\alpha\beta}&-&4\sum_{j<k}
\Re\left(\tilde{X}^{\alpha\beta}_j\tilde{X}^{\alpha\beta\ast}_k
\right)\sin^2\left({\Delta \tilde{E}_{jk}L \over 2}\right)\nonumber\\
&+&2\sum_{j<k}
\Im\left(\tilde{X}^{\alpha\beta}_j\tilde{X}^{\alpha\beta\ast}_k
\right)\sin\left(\Delta \tilde{E}_{jk}L\right),
\nonumber
\end{eqnarray}
where $\Delta\tilde{E}_{jk}\equiv\tilde{E}_j-\tilde{E}_k$.

\subsection{Parameter degeneracy}

Naively one might assume that with two general unknowns, 
($\theta_{13}, \delta$), 
and one discrete unknown, (the sign of $\Delta m_{32}^2$), and two 
measurements of the oscillation 
probability at a neutrino factory (i.e. $\nu_e \to \nu_\mu$ and 
$\bar\nu_e \to \bar\nu_\mu$ ) it would be straightforward to extract
the mixing parameters.  However, 
since the work \cite{Burguet-Castell:2001ez},
it has been realized that
there is in general more than one solution for
the set of oscillation parameters,
even if we know both $P(\nu_\mu\rightarrow\nu_e)$
and $P(\bar\nu_\mu\rightarrow\bar\nu_e)$.
Reference \cite{Burguet-Castell:2001ez} shows that there
are intrinsically two sets of solutions ($\theta_{13}$, $\delta$)
and ($\theta'_{13}$, $\delta'$).
Reference \cite{Minakata:2001qm} describes 
the degeneracy under
$\Delta m^2_{32}\leftrightarrow -\Delta m^2_{32}$, and
Reference \cite{Fogli:1996pv,Barger:2001yr} found the ambiguity
under $\theta_{23}\leftrightarrow\pi/2-\theta_{23}$.
Thus there is an eight-fold degeneracy \cite{Barger:2001yr},
and the resolution of this degeneracy was one of the main
subjects at Nufact'02.
The degeneracy in $\theta_{23}\leftrightarrow\pi/2-\theta_{23}$
is difficult to resolve as long as we look at
only channels $\nu_\mu\leftrightarrow\nu_e$,
$\bar\nu_\mu\leftrightarrow\bar\nu_e$
and $\nu_\mu\rightarrow\nu_\mu$, since the first two
create the degeneracy while the third gives
the information only on $\sin^22\theta_{23}$.
As was pointed out in \cite{Barger:2001yr},
the best way is probably to look at the channel
$\nu_e\rightarrow\nu_\tau$ whose oscillation
probability in vacuum is given by
$\cos^2\theta_{23}\sin^22\theta_{13}\sin\left(\Delta m^2_{32}L/4E\right)$,
which is compared with
$P(\nu_e\rightarrow\nu_\mu)=
\sin^2\theta_{23}\sin^22\theta_{13}\sin\left(\Delta m^2_{32}L/4E\right)$.

To determine $\theta_{13}$,
Minakata \cite{Minakata:2001qm} proposed to tune the neutrino
energy at the oscillation maximum ($\Delta m^2_{32}L/4E=\pi/2$)
so that the ellipse in the $P(\nu_\mu\rightarrow\nu_e)$ --
$P(\bar{\nu}_\mu\rightarrow\bar{\nu}_e)$ plane that results from 
letting $\delta$ run from 0 to $2\pi$ 
collapses to a line.\footnote{The idea to tune the
neutrino energy at the oscillation maximum was
given earlier in \cite{Barger:2001qs}.}
Then the ambiguity in $\theta_{13}$ is reduced.

Whisnant \cite{whisnant} discussed systematically whether or not a set
of measurements in neutrino oscillation appearance experiments can
completely resolve the ambiguities.  He showed that experiments with
neutrinos and antineutrinos at two different energies, which give
the four independent measurements, can in principle resolve all of the
parameter degeneracies if $\sin^22\theta_{13} \geq 0.002$.

Mena \cite{mena} showed that by combining a superbeam and a neutrino
factory, it is possible to reduce the intrinsic,
sign($\Delta m_{13}^2$) and $\theta_{23}$ degeneracies.  She pointed
out that the short neutrino factory baseline (e.g. $L=732$ km) becomes
an interesting distance to combine with a 
superbeam experiment for values of
$\theta_{13}$ near its present limit.  For smaller $\theta_{13}$, an
intermediate neutrino factory baseline of $O$(3000km) is still required.

Meloni \cite{meloni} showed how the $\nu_e \to
\nu_\mu$ (``golden'') and $\nu_e \to \nu_\tau$ (``silver'')
transitions observed at an OPERA-like 2 Kton lead-emulsion detector at
$L = 732$ Km, in combination with the $\nu_e \to \nu_\mu$ transitions
observed at a 40 Kton magnetized iron detector with a baseline of $L =
3000$ Km, strongly reduce the intrinsic $(\theta_{13}, \delta)$
ambiguity.  He also showed how a moderate increase in the OPERA-like
detector mass (4 Kton instead of 2 Kton) completely eliminates
the degenerate regions even for small values of $\theta_{13}$.

Huber \cite{Huber:2002uy} compared the physics potential of planned
superbeams with the one of neutrino factories.  He showed that
JHF--SuperK (0.75MW, 110Kt$\cdot$yr),
JHF--HyperK (4MW, 8Mt$\cdot$yr) and NuFact--I (0.75MW, 50Kt$\cdot$yr),
NuFact--II (4MW, 400Kt$\cdot$yr)
can explore $\sin^22\theta_{13}$ down to $\simeq 2\cdot10^{-2}$,
$\simeq 10^{-3}$, $\simeq 5\cdot10^{-4}$, respectively, while
only JHF--HyperK and NuFact--II can probe the CP phase
for $\sin^22\theta_{13}\gtrsim$several$\times10^{-3}$, $10^{-4}$,
respectively.

A previous study showed that two simultaneous baselines for 
a neutrino factory could be used to resolve degeneracies 
\cite{Burguet-Castell:2001ez} 
where something like a combination of 
3000km plus either 700km or 7000km 
baselines might be optimal.  For one neutrino factory
to access two baselines, a triangular or bow-tie ring would be 
necessary.  Depending on what the first generation of neutrino 
superbeams measures, the argument for making two simultaneous 
baselines may be less compelling.  

\subsection{Matter effect}
In very long baseline experiments, it is important to
estimate the magnitude of the matter effect.
Unlike in shorter or medium long baseline experiments,
in which neutrinos go through only the crust or the upper mantle
of the Earth, a constant density approximation is not good
in very long baseline experiments.

By using the method of a Fourier series expansion of the matter
profile, Ota \cite{ota} showed that neglecting the matter profile effect
leads to an extra uncertainty ($\sim$ 5\%) on the average density for
5,000km $\lesssim$ $L$ $\lesssim$ 10,000km.  He proposed to treat the
first coefficient in the Fourier series as a fitting parameter.

Winter \cite{winter} compared three different methods
(single perturbation, random fluctuations, and
the measured mean)
to evaluate uncertainties in the
Earth's matter density profile.
He gave advantages and disadvantages for each method
and he concluded that
matter density uncertainties probably would not be the bottleneck of
the statistical analysis of a planned experiment, while
he claimed that somewhat more
effort should be spent on improving the results from geophysics.

\subsection{Exotic scenarios}

Although the LSND result \cite{Athanassopoulos:1996jb} is not yet
confirmed, we can still try to explain the LSND result in at least three  
different ways.  The first way is to assume a fourth mass eigenstate, 
which implies one sterile neutrino (with a (2+2) or (3+1) mass scheme, 
depending on the mass
pattern).  The LSND signature is then simply the result of 
three independent mass squared differences.  
Valle \cite{valle} gave the results of his recent
analyses on the solar and atmospheric neutrino data for the
(2+2)--scheme, and it is strongly disfavored.  The (3+1)--scheme is
allowed only at 99\%CL.  A second possibility is to give up
CPT invariance \cite{Murayama:2000hm} and to take full advantage
of the difference between neutrinos and antineutrinos.  This
possibility would be excluded if KamLAND observes the deficit.
The third idea \cite{babu} is to assume a lepton number violating interaction
$\mu^+\rightarrow e^+ + \bar{\nu}_e + \bar{\nu}_\alpha$
($\alpha=e, \mu, \tau$).  Interestingly, this scenario predicts that
the MiniBooNE experiment will see no signal.
There are other exotic scenarios which can be tested at neutrino factories.

Wong \cite{wong} considered the situation
where the equivalence principle is violated
in neutrino oscillations with three flavors.
The $3\times3$ matrix which describes the propagation of
neutrinos is
\begin{eqnarray}
\hspace*{-20mm}
H=U\mbox{\rm diag}(E_1,E_2,E_3)U^{-1}
+\mbox{\rm diag}(\sqrt{2}G_FN_e,0,0)
+2E|\phi|U_G\mbox{\rm diag}(\gamma_1,\gamma_2,\gamma_3)U_G^{-1},
\nonumber
\end{eqnarray}
where $\gamma_j$ is the relative gravitational coupling
constants,
Violation of the Equivalence Principle (VEP) implies
$\gamma_j\ne 1$, $U_G$ stands for the mixing matrix
between the gravitational and flavor eigenstates, and
$\phi$ is the gravitational potential.
Assuming $|\Delta\gamma|\equiv|\gamma_3-\gamma_2|\gg|\gamma_2-\gamma_1|$
for simplicity, she gets the following ratio of
T violation $\Delta \tilde{P}_T$ with VEP to
that $\Delta P_T$ without VEP
\begin{eqnarray}
{\Delta \tilde{P}_T \over \Delta P_T}=
1-{4E^2 \over \Delta m^2_{21}}
{\sin2\theta_{\rm atm} \over \sin2\theta_\odot}
\left|\phi\right|\Delta\gamma\sin2\varphi,
\nonumber
\end{eqnarray}
where $\varphi$ is the mixing angle between the first and the third
eigenstates in $U_G$ and other mixings in $U_G$ are assumed to be
zero for simplicity.
She showed that T violation $\Delta \tilde{P}_T$ with VEP
is deviated from $\Delta P_T$ without VEP
if the violation $|\Delta\gamma||\phi|$ is large
and therefore may be measured in long baseline experiments.

Sato \cite{sato} discussed the measurement of new physics
which is described by an interaction such as
$g(\bar{e}\gamma_{{}_\lambda}\mu)(\bar{\nu}_\mu
\gamma^{{}_{\tiny\lambda}}\nu_\beta)$
($\beta\ne e$) in long baseline neutrino oscillation experiments.
Through the neutrino oscillation, 
the probability to detect the new physics effects such as 
flavor violation is 
enhanced by the interference with the weak interaction.
Assuming a neutrino factory and an upgraded conventional beam, 
he examined the possibility to observe new physics numerically  
and pointed out that it is possible to search for new interactions 
using some channels, for example
$\nu_{\mu} \rightarrow \nu_{\mu}$, in these experiments.
He also showed that there is a chance to observe a
flavor-changing effect in $\tau$ appearance experiments
such as ICARUS and OPERA.

Campanelli \cite{campanelli} discussed
a new flavor changing interaction which
gives the contribution $(\epsilon_{\alpha\beta})$
($\alpha\beta=e, \mu, \tau$) to the $3\times 3$
matrix in the flavor basis.
He showed that the new physics effect becomes
important in the channels $\nu_e\rightarrow\nu_\alpha$
($\alpha=\mu, \tau$)
at sufficiently high neutrino energies,
since the contribution from the conventional physics
is suppressed by $\theta_{13}$ and $1/E_\nu$.
Therefore it may be possible to observe the new physics effect
particularly by looking at the energy spectrum which would
be different in the presence of new physics.

\section{Oscilation Measurements} 

%


It is clear from the discussion above that to 
measure the 
oscillation 
parameters and to 
probe
exotic
scenarios,
we will need more than one measurement of oscillations 
to understand phenomena visible at the atmospheric $\delta_m^2$.
Furthermore, there 
is much than can be learned at high intensity, 
conventional 
(that is, meson decay)
beams.  If a signal is seen in a conventional 
beam, that information may still be an important constraint even in the 
advent of a neutrino factory
because of the different energies of the beams.  
The challenge with doing any of the measurements described above 
in conventional neutrino 
beams is  
%
that the $\nu_\mu\to\nu_e$ transition probability is limited by the 
CHOOZ non-observation of $\nu_\mu$ disappearance to be small, less than
$5\%$ in the absence of enhancement from CP violation or matter effects.
%
Intrinsic $\nu_e$ backgrounds in conventional beams at the 
per cent level, due primarily to the decay chain 
$\pi^+ \to \nu_\mu \mu^+, \mu^+ \to e^+ \nu_e \bar\nu_\mu$,
and significant background processes present if neutrino energies are
above a few hundred MeV from neutral current events mimicking
electron signals will make these measurements difficult.

For a given detector mass, one gets the best sensitivity for 
seeing a non-zero
value of 
the muon to electron neutrino transition probability 
by placing an experiment near or slightly below
the first oscillation maximum ($1.27\Delta m^2 L/E \approx \pi/2$).   
In order to see evidence for matter effects, and thereby deduce the mass 
hierarchy, however, one needs to go to baselines on the order of 
1000km, and therefore use neutrino energies well above 1GeV.  Finally, to look 
for CP violation, one might still run at the first oscillation peak, as 
described by Minakata\cite{Minakata:2001qm}, but it has also 
been argued that one should consider running an experiment at 
the second or third oscillation maximum \cite{marciano}, 
since the CP-violating contribution to the transition grows with $L/E$.

To get around the problem of backgrounds, several different strategies 
have been proposed.  For example, the JHF to SuperK, the 
NUMI Off-axis, and the CNGT proposals all take advantage of 
the two-body decay kinematics 
of the pion, and put the far detector slightly off the main axis of the 
neutrino beamline 
\cite{jhf,autin,kopp}.
This trick, first suggested in BNL proposal 
E889 \cite{bnl889}, 
allows one to maximize the $\nu_\mu$ flux in a very narrow energy 
region, reduces the flux at higher neutrino energies, and does 
not alter the flux from three-body decays, which are the
source of intrinsic $\nu_e$'s in the beam.  
This results in an improved signal to background ratio at the peak beam energy
since the background from high $E_\nu$ neutral current interactions that fake
electron charged current events at this peak energy is reduced.
However, since
the $\nu_\mu$ spectrum is significantly narrower than the shape of the 
appearance probability, one suffers from statistics.  This must be 
mitigated by the use of massive detectors and very high proton beam power.  

Another way to reduce the effect of backgrounds is to consider a 
distance 
of thousands of kilometers 
where the matter effects will significantly amplify either the 
neutrino or the antineutrino probability, so that even for a small value
of $\sin^2 2\theta_{13}$ a large probability would be expected.  This is 
the strategy of the Brookhaven to NUSL proposal, as described by 
S.Kahn \cite{skahn}.  However, to go such large
distances, even with a wide band beam, requires 
a ``megaton class'' detector.
By combining a large distance and a broad band neutrino energy 
spectrum, this proposal can see not only the first but the second oscillation 
maximum, and as such can have more constraints on the oscillation parameters.  

A third way to reduce the backgrounds inherent in a conventional 
beam is to go to 
neutrino energies of a few hundred MeV
and 
to
construct only 
a short decay tunnel so that few muons decay to produce $\nu_e$'s, 
as described by M.Mezzetto.  
The intrinsic $\nu_e$ contamination of the 
proposed CERN
SPL beam 
is at the parts per mil level,
and at 
this neutrino energy 
the few neutral pions that do get produced are easy to 
distinguish from electrons
since there is a large angle between the two decay photons.   
However in the few hundred MeV energy range, the cross-section varies
quadratically with energy, suppressing the rate,
so again 
a megaton-class
detector must be used.  

Finally, a novel technique for creating a beam with extremely 
low backgrounds is to accelerate a beam of radioactive ions in a long 
storage ring, and let them decay to electron neutrinos.  The intrinsic 
backgrounds here are just as low as that of a neutrino factory, and in fact
the beam is really only $\nu_e$ or $\bar\nu_e$ depending on the ion in the
ring, with contaminations from any other flavor at less than the 
$10^{-4}$ level.  The resulting neutrino beam would be at a very low 
neutrino energy, and a water {\cyr Qerenkov} 
\footnote{Here we use the correct Cyrillic spelling and not the
often seen used ethnically inaccurate version with the \v{C}ech accent}
detector located in the Frejus 
tunnel could be used for both this 
``beta beam'' concept and the CERN SPL neutrino beam.  This concept 
was presented in the plenary session by P.~Zucchelli and in a 
parallel session by M.~Mezzetto.  

Table \ref{tab:exps} gives a summary of the different neutrino experiments
that are being considered worldwide, and the different strategies being
adopted.  The experiments listed below are at very different stages of 
completion, as well:  the JHF to SuperKamiokande uses an existing 
$50$~kTon detector, but will require a new beamline to be built 
at the Japanese Hadron Facility.  The beamline is being designed not only 
to handle both a first stage experiment with a 1MW proton source, but also 
an upgraded stage with a 4MW proton source, and a 
megaton-class
water {\cyr Qerenkov} detector \cite{ichikawa}.  
The NUMI Off-axis experiment would use the (almost) 
existing NUMI beamline, but a new 
$\sim30$~kTon detector
would need to be built
somewhere in northern Minnesota or perhaps even in Canada \cite{kopp}.  
The CNGT project would require instrumenting a large body of water, perhaps
in the gulf of Taranto for a detector, and the CNGS beamline would have to 
be tuned at a lower energy which would preclude measuring 
$\tau$ appearance \cite{autin} at Gran Sasso.  
Finally, the other programs (CERN SPL and the BNL 
proposal) require both new beamlines and new detectors 
\cite{skahn,mezzetto}.  

\begin{table} 
\label{tab:exps} 
\begin{tabular}{lccccccc} 
Program & $<E_\nu>$ & L  &  Proton  &  Detector & Wide or & Which &  Matter \\
Name    & GeV       & km &  Power   &    Mass   &  Narrow & max   &  Effect \\
        &           &    &  (MW)    &   (kton)  & Band    & ?     &  Size \\
\hline 
JHF to SK$^a$   & 0.77  & 295 & 0.75  &  22.5  & Narrow    &   1  &  tiny \\ 
NUMI OA $^b$    & 2     & $\simeq 730$ & 0.4 & 20 & Narrow &   1  &  big \\ 
BNL to NY $^c$  &  1    &     &   1     &  500    &  Wide  &   1  &  tiny \\ 
BNL to NUSL $^c$  &  1    & 2600 &   1  &  500    &  Wide  &   1  &  huge \\ 
SJHF to HK $^a$  &  0.77 & 295  & 4    & 1000 &   Narrow &   1  & tiny \\ 
SNUMI OA  $^b$ & 2  & $\simeq 1000?$ & 2 & 100 & Narrow &   1  &  bigger \\ 
SPL  $^d$    & 0.27  &  130  &  4    &  500  &  Wide  &   $<1$   & none \\ 
CNGT   $^e$    &  0.8 & 1200 &  $\sim 0.5^{ee}$      &  1000    &  Narrow   &  1   &  big \\
$\beta$ beams$^d$ &  0.13  &  130  &  n/a   &  500  &  wide &  1 & none \\
$\nu$ Factory$^f$ & 10-30 & 3000 & 4    &  50 &   Wide & $<1$ &  huge \\ 
\end{tabular} 
\caption{Vital characteristics of future possible neutrino experiments. 
$^{ee}$ Proton power of CNGT assumed to be approximately that of CNGS.  
Speakers in this workshop were: 
$^a$A.~Ichikawa, $^b$S.~Kopp, $^c$S.~Kahn, 
$^d$M.~Mezzetto,$^e$B.~Autin, $^f$S.~Geer.  } 
\end{table} 

\subsection{Far Detector Strategies} 

     For most of the proposals listed in table \ref{tab:exps}, the 
detector of choice is a 
water {\cyr Qerenkov} device--either the SuperKamiokande
detector, or a new larger version, which would also presumably be 
instrumented well enough to search for proton decay and supernovae 
neutrinos.  
At energies at or below $1$~GeV,
the SuperKamiokande detector is
able to keep detector misidentification at 
a level comparable to or below the intrinsic electron neutrino 
backgrounds in the JHF off-axis beamline.  

     For a neutrino factory, the energies are 
$10$s of GeV,
and because 
charge identification is required, there is 
an appropriate and proven candidate detector technology,
a steel-scintillator detector like the one being built for the MINOS 
experiment \cite{golden}.  
For the $\nu_e\to \nu_\tau$ measurements that were described
by Sato and Meloni, an OPERA-like detector, even one that was twice the 
mass of the current design, could provide very interesting measurements.  

     A detector technology which in principle could work for any of 
the conventional beamlines, is a liquid argon TPC, such as that 
being built by the 
ICARUS collaboration.  Its very precise imaging capabilities would mean
very good background rejection, while maintaining high efficiency, as 
described by D.~Cline.  Because the backgrounds from neutral current 
events can be eliminated, and the signal efficiency remains high, the 
reach per kton of a liquid argon detector can be as much as 3 times 
that of other technologies.  Therefore, one would require in principle 
a much smaller detector mass than those listed in the table above.  
Although using the 600 Ton module size of ICARUS would be prohibitively 
expensive even for a 6kton fiducial mass, 
K.~McDonald showed that for large enough single vessels, 
and large enough wire plane separation, this may become a very economical 
technology.  Certainly the technology to make extremely large
(order 100kton) cryogenic vessels is well-understood thanks to the natural
gas companies, and the biggest challenge will likely be elsewhere.  

     For the NUMI Off Axis beamline, another detector technology being 
considered is fine-grained calorimetry, as described in a talk by A.~Para.  
Fine grained calorimeters have a long history in the field of neutrino 
physics, and are well-understood and well-simulated.  By making a low-Z 
calorimeter which was segmented every third of a radiation length, one 
could build a detector with reasonably high efficiencies, which would 
be able to reduce the neutral current background in the NUMI Off-Axis 
beamline to below that of the intrinsic $\nu_e$ contamination, which is 
itself about $0.4\%$ under the peak energy.  The challenge here will be 
to design a device that is economical to construct
and realizes these large background reductions.  

\subsection{Measurement Issues in Few GeV Beams} 

     As can be seen by the large number of proposals being considered, 
it is clear that one or more ``neutrino superbeams'' are likely to 
provide measurements before the advent of a neutrino factory.  What's more,
these neutrino superbeams are likely to be between a few hundred MeV region, 
such as the CERN SPL proposal, to the few (up to 6?) GeV region as in 
the BNL proposal.  The experimental challenges associated with measurements
with these beamlines are extremely different from those associated with 
those from a neutrino factory.  

     In the absence of oscillations, 
the number of events at a detector is simply the product of the
neutrino flux times the cross section times the detector efficiency.  
Uncertainties in any of these can 
degrade the ability to use precise measurements of multiple transition 
probabilities to extract oscillation parameters,
and 
in fact all of them may have 
complicated
energy dependence, 
in particular in the 
sub to few 
GeV region.  
To address these issues, we convened a
joint session with 
the neutrino scattering working group.

\subsubsection{Cross Sections} 

At very low energies, the cross-section is dominated by elastic or 
quasi-elastic processes which are few body reactions and relatively
easy to model, at least in a free nucleon target.
At very high energies, the cross-section is dominated by deep inelastic 
scattering, where complex multi-body final states
can be effectively modeled with small corrections to a dominant picture of
neutrino-quark scattering with an outgoing ``jet'' created by recoil hadronic
system.  However,
the neutrino energies are on the order of the 
nucleon mass
the second portion of the cross-section becomes dominated by intermediate
resonant hadronic states with discrete masses, which complicate the picture 
significantly.  In addition, the few body nature of the final state requires
modeling exclusive reactions in order to correctly predict signal and
background efficiencies.
%

%
Quasi-elastic cross-sections can be predicted very accurately, although the
effects of final-state interactions in nuclei give $10$--$20$\% uncertainties
in observed rates.  Deep inelastic cross-sections are known to a few percent
in the high energy limit.  However, in the ``resonance'' region the rates for
exclusive final states typically have $50$--$100$\% uncertainties, and
often differential distributions are completely unmeasured as discussed
by Casper at this workshop, and as shown in Reference \cite{skat}.
Bodek has been looking at a way to extend deep 
inelastic scattering predictions for cross sections to the resonance 
regime using global fits of many different data sets, and has shown that 
for charged lepton scattering, this treatment results in an accurate prediction
of the mean cross section in the resonance region.  By then using the data 
from charged leptons in the resonance region, it is hoped that this 
treatment can result in correct 
modeling of resonances in neutrino scattering at 
low energies \cite{bodek}.  

\subsubsection{Flux Predictions} 

To first order, flux predictions will be made in these upcoming
superbeam experiments by simply measuring the neutrino event
distributions as a function of energy in a near detector and then by
using a beamline simulation to extrapolate from near to far.  However,
this is complicated by the fact that the neutrino source is usually so
close to the near detector that the source does not appear point-like.
Also, if the neutrinos are from 3-body or non-pion decays, (i.e. the
$\nu_e$'s, which are from three-body decays of the muon and kaon) the
near-far extrapolation is different.

Several different strategies for making flux predictions were discussed 
in this workshop.  One very straightforward strategy in 
principle is to simply measure directly (or indirectly) 
the distributions of the secondaries 
that are creating the tertiary neutrinos in the first place.  Kobayashi 
discussed the data from the pion monitor used by K2K, which gives the 
angular distribution of pions at a given radius.  
The pion angular distribution, combined with the near detector measurements, 
gives a systematic uncertainty on the flux extrapolation 
to the far detector of 5\% \cite{kobayashi}.    
Coney discussed how MiniBooNE would be using muon counters located at the 
end of a small pipe set at a large angle with respect 
to the decay pipe at MiniBooNE.
Most 
of the muons which were accepted at the end of that
pipe
are products of
high angle kaon decays, 
and 
by measuring the energies of those muons, one could get 
a good constraint (estimated uncertainty goal of 10\%) 
on the ratio of pion to kaon decays in the beamline.  
This constraint is important for limiting the uncertainty in the kaon-induced
$\nu_e$ background in MiniBooNE.  

Another strategy for extrapolating from the near to the far detector was
outlined by Szleper, using as an example the both the on and off-axis 
fluxes expected at the NUMI 
beamline.  He considered, among other effects, the 
uncertainties in near to far extrapolation due to hadron 
production uncertainties.  Because the 
bulk of the $\nu_\mu$'s expected in both the near and far detectors are 
from pion decays, and the low energy $\nu_e$'s from muon decays, 
where the muons were also from pion
decays, Szleper showed how there is a straightforward correlation between
what is measured in a near on-axis detector and far detectors, both on 
and off axis.  The standard technique is simply to take the near detector
energy spectrum and use the beamline simulation to create a ``far over near
ratio'' for each energy bin, and then multiply by that ratio.  
The new technique developed by Szleper and Para, however, 
involves a ``matrix approach'' where for every 
neutrino event at the near on-axis detector there is a spectrum of events
at the far detector, and those events are not be all reconstructed with the 
same energy as the event in the near detector \cite{paraszleper}.  
For the $\nu_\mu$'s coming from the focused pions, the two techniques 
result in similar hadron production uncertainties of 1-2\%, but for neutrinos
for higher energy pions, those just underfocused, the differences are 
reduced from 15\% (using the traditional approach) to 5\%. 
Para 
offered a wager of
a bottle of good sake 
during the conference that if K2K were to use this technique their 
far detector flux would be predicted to 1\%, rather than to the 5\% 
uncertainty currently assigned \cite{kobayashi}.  
The error has yet to be reduced and no sake has yet changed hands.

Of course at a muon storage ring, the issues of flux prediction are quite 
different because the beam and decay kinematics as well known.  As evidence
of this point, Broncano showed that the ${\cal O}(\alpha_{EM})$ corrections
to the neutrino distributions were below the $0.1\%$ level \cite{Broncano}.
Therefore, it is likely that a neutrino factory will be in the enviable
position of having the primary challenge to knowing the flux be
knowledge of the phase space distribution and number of muons in the beam.

\subsubsection{Near Detector Strategies} 

Near Detectors play a very different role depending on whether or 
not one is discussing disappearance or appearance measurements.  
In the case of disappearance measurements, where it is assumed the 
oscillation probability will be large and the backgrounds to the 
signal small, the near detector is used to make an accurate 
prediction of the $\nu_\mu$ flux.  In particular for a wide band 
neutrino beam, the energy dependence of both the flux and the 
detector response must be well understood.  For appearance measurements, 
however, one wants to use a near detector to measure the backgrounds, 
and in particular how the backgrounds enter as a function of the 
neutrino energy, and the measured (visible) energy.  For both kinds 
of measurements there are conflicting requirements for the near detector: 
one would like a detector that resembles the far detector as much as 
possible, but at the same time one would like a near detector that could 
make detailed studies of neutrino interactions in order to best model 
the far detector response.  

     One solution to this near detector problem is to have a suite of near
detectors, as has been done very successfully by the K2K experiment, 
reported by Yoshida.  While they have a 1kTon water {\cyr Qerenkov} 
device for a 
near to far detector ratio where the cross sections cancel, they can also 
compare events in the near water {\cyr Qerenkov} detector with 
those seen in their 
fine-grained scintillating fiber detector.  The fine-grained detector can 
see the recoil proton in the neutrino interaction, and K2K can use this
to deduce the energy resolution and acceptance of their far detector.  
Furthermore, there is also a large muon chamber which can see the 
profile of the neutrino beam and verify that indeed the beam is pointing
towards the far detector.  

The success of this near detector suite is 
evidenced by the fact that very similar plans are in place for an off-axis 
near detector suite for the JHF to SuperKamiokande experiment, as described
by Kajita.  There will be 
near detectors at 280m to study cross sections and verify the position of 
the neutrino beam, and a ``less near'' detector at 2km consisting of a 
water {\cyr Qerenkov} device to make the most accurate far/near prediction
\cite{jhf,kajita}.  

     Because of the difference in energies and detector technologies, the 
MINOS experiment can use one near detector which will be used to not only 
measure the flux and event rates (using both quasielastic and 
non-quasielastic events, respectively), but which can 
also verify the direction of 
the neutrino beam (during special high energy runs). As described by 
Morfin, MINOS is also considering adding a specially-modified 
cross section detector in front of the near detector, which can study 
neutrino interactions more precisely, and also can study the Z dependence
of neutrino cross sections by having planes of various materials:  carbon, 
steel, and tungsten \cite{Morfin}.  

     For a NUMI off-axis experiment, however, one would like to study 
the detector-related backgrounds with an off-axis detector.  This is because 
the on-axis beam spectrum is very different from the off-axis spectrum, 
and also because 
the visible energy in the background events is often far from the 
initial neutrino energy.  One definite requirement for an off-axis 
experiment would be an off-axis near detector which mimics a far 
detector, but there have also been thoughts about also including 
a more fine-grained near detector, \` a la K2K.  Bodek described a
possible new detector for cross section studies that were 
relevant for a NUMI off-axis experiment, as well as a possible site for 
such a detector in the NUMI beamline \cite{bodek}.  

\section{Conclusions} 
From this workshop, it is clear that conventional neutrino beams will
occupy the foreground of oscillation measurements until perhaps 2010. 
Then, a crucial
decision will have to be made in choosing between hyper conventional
experiments on one hand combining multi MegaWatt proton beams combined
with MegaTon detectors, and 
neutrino factories on the other hand whose design should
be mature by that time.


\end{document}